\documentclass[twocolumn]{aastex631}
\usepackage{amsmath}
\usepackage{svg}
\usepackage{xcolor}

\def\mailto#1{\href{mailto:#1}{#1}}

\shorttitle{Kernel Selection for Gaussian Process}
\shortauthors{Zhang et al.}
\graphicspath{{./}{figures/}}

\begin{document}

\title{Kernel Selection for Gaussian Process in Cosmology: with Approximate Bayesian Computation Rejection and Nested Sampling}

\author[0000-0002-5595-0447]{Hao Zhang}
\affiliation{Institute for Frontiers in Astronomy and Astrophysics, Beijing Normal University, Beijing 102206, China}
\affiliation{Department of Physics, Beijing Normal University, 
Beijing 100875, China}
\affiliation{Key Laboratory for Particle Astrophysics, Institute of High Energy Physics, CAS 19B Yuquan Road, Beijing 100049, China}
\affiliation{School of Astronomy and Space Science, University of Chinese Academy of Sciences, 19A Yuquan Road, Beijing 100049, China}

\email{haozhang@ihep.ac.cn}


\author[0000-0002-8429-7088]{Yu-Chen Wang}
\affiliation{Department of Astronomy, School of Physics, Peking University, 
Beijing 100871, China}
\affiliation{Kavli Institute for Astronomy and Astrophysics, Peking University, Beijing 100871, China}

\author[0000-0002-3363-9965]{Tong-Jie Zhang}
\affiliation{Institute for Frontiers in Astronomy and Astrophysics, Beijing Normal University, Beijing 102206, China}
\affiliation{Department of Astronomy, Beijing Normal University, 
Beijing 100875, China}
\affiliation{Institute for Astronomical Science, Dezhou University, 
Dezhou 253023, China}
\email{tjzhang@bnu.edu.cn}

\author{Tingting Zhang}
\affiliation{College of Command and Control Engineering, PLA Army Engineering University, Nanjing 210000, China; \mailto{101101964@seu.edu.cn}}



\begin{abstract}

Gaussian Process (GP) has gained much attention in cosmology due to its ability to reconstruct cosmological data in a model-independent manner. In this study, we compare two methods for GP kernel selection: Approximate Bayesian Computation (ABC) Rejection and nested sampling. We analyze three types of data: cosmic Chronometer data (CC), Type Ia Supernovae (SNIa), and Gamma Ray Burst (GRB), using five kernel functions. To evaluate the differences between kernel functions, we assess the strength of evidence using Bayes factors. Our results show that, for ABC Rejection, the Matérn kernel with $\nu$=5/2 (M52 kernel) outperformes the commonly used Radial Basis Function (RBF) kernel in approximating all three datasets. Bayes factors indicate that the M52 kernel typically supports the observed data better than the RBF kernel, but with no clear advantage over other alternatives. However, nested sampling gives different results, with the M52 kernel losing its advantage. Nevertheless, Bayes factors indicate no significant dependence of the data on each kernel. 

\end{abstract}



\section{Introduction} \label{sec:intro}
Given the widespread use of Gaussian Process in cosmology, it is crucial to address various related issues. Cosmologists often use Gaussian Process to reconstruct cosmological data to anticipate specific cosmological parameters at a given redshift and constrain these parameters through prediction (\cite{gomez2018h0}; \cite{dhawan2021non}). Gaussian Process can also be utilized to address the Hubble Tension\citep{mehrabi2020does}, constrain cosmological parameters in a non-parametric way\citep{mukherjee2021non}, and deal with relevant issues in the realm of dark energy \citep{seikel2012reconstruction}, as it can reconstruct data in a model-independent manner. As Gaussian Process is widely used, the specifics of its calculation, including hyperparameters, quantity of hyperparameters, and choice of kernels, can impact the reconstruction of cosmological data and our predictions.

The choice of hyperparameters, the quantity of hyperparameters, and the choice of kernels can all affect the results of the Gaussian Process. Earlier studies have addressed many of these issues. For instance, \cite{sun2021influence} studied the influence of the upper bound and lower bound of the two hyperparameters in the radial basis kernel function (RBF). There were also some articles about the choice of kernels. However, these studies, such as \cite{PhysRevD.85.123530} and \cite{hwang2023use}, did not provide sufficient details on the choice of kernel functions and only spent a relatively small portion of their space on the issue without a clear selection scheme. Others, such as \cite{o_colgain_elucidating_2021}, did not cover a sufficient variety of kernel functions.Therefore, our research aims to evaluate which kernel outperforms others for various datasets when various distance functions are used to assess the effect of reconstruction and the degree to which each kernel is supported by observed data.

To accomplish our objectives, we use Approximate Bayesian Computation (ABC) Rejection combined with three distinct distance functions to evaluate the performance of five kernels against three different cosmological datasets, including cosmic chronometer data (CC), Type Ia Supernovae (SNIa), and Gamma Ray Burst data (GRB). Additionally, we calculate the Bayes factor between every two kernels based on the posterior distribution provided by the ABC Rejection to show the subtle differences in support strength of our datasets for the five kernels we test. The Bayes factor $K$ is a powerful tool for quantitatively comparing the strength of evidence for two models from observed data given the prior distribution of two different models. This allows us to determine exactly how superior the ideal kernel function is to alternative kernels.

This paper's structure is as follows. In Section \ref{sec:methods}, we introduce Gaussian Process (Section \ref{GP}) and other mathematical methods, such as Approximate Bayesian Computation (ABC) Rejection (Section \ref{ABC}), distance function (Section \ref{dis_f}), Bayes factor (Section \ref{Bayesfactor}), and nested sampling (Section \ref{nestsample}), briefly. In Section \ref{sec:results}, we describe the results of calculations using the mathematical methods mentioned for the cosmological datasets we use, respectively, involving ABC Rejection, Bayes factor, and nested sampling. In the last section, Section \ref{sec:conclusion}, we summarize our results and discuss the caution that scientists should exercise in kernel selection when using Gaussian Process to reconstruct data.

\section{Methods} \label{sec:methods}
\subsection{Gaussian Process(GP)}\label{GP}
The Gaussian Process (GP) is a powerful tool for modeling functions in a stochastic statistical process. It represents a collection of random variables with a joint Gaussian distribution from the perspective of function-space \citep{williams2006gaussian}. The observations in GP are presented in a continuous domain, where a normally distributed random variable is connected to each point in the input space. By using GP, we can predict the value of a function at a new point without assuming the form of the function, based on the observed data points. Specifically, a Gaussian random variable at a reconstructed point $z$ represents the expected value for GP.

To forecast the function value at $z$, only the covariance function and mean function of GP are needed. Let $\boldsymbol{Z}=\{z_i\}$ be a vector of known redshift data points. The crucial step in obtaining the mean and covariance function is calculating the covariance matrix $\emph{K\textbf{(Z,Z)}}$, which is determined by the observed data, hyperparameters, and kernel function $k(z_i,z_j)$ for arbitrarily chosen reconstructed points $z_i$ and $z_j$. The covariance matrix $\emph{K\textbf{(Z,Z)}}$ displays the relationship between different reconstructed points, indicating that function values at various places in GP are not independent of one another. The kernel function returns a single element, $K_{ij}$, of the covariance matrix $\emph{K\textbf{(Z,Z)}}$ \citep{seikel2012reconstruction}:
\begin{equation}
    [\emph{K\textbf{(Z,Z)}}]_{ij}=k(z_i,z_j).
\end{equation}

The primary goal of GP is to determine the mean and variance of the vector $\boldsymbol{f^*}=\{f(z_i^*)\}$ containing the reconstructed function values at prediction points. In order to obtain these values, the mean $\overline{\boldsymbol{f^*}}$ and covariance matrix cov($\boldsymbol{f^*}$) must be calculated using the approach outlined in \cite{williams2006gaussian}. Specifically, Equation \eqref{mean&cov} provides the expressions for these quantities:

\begin{align}\label{mean&cov}
    \overline{\boldsymbol{f^{*}}}&=K(\boldsymbol{Z^{*}},\boldsymbol{Z})[K(\boldsymbol{Z},\boldsymbol{Z})+C]^{-1}\boldsymbol{y},\\
    \mathrm{cov}(\boldsymbol{f^{*}})&=K(\boldsymbol{Z^{*}},\boldsymbol{Z^{*}})\nonumber\\ &-K(\boldsymbol{Z^{*}},\boldsymbol{Z})[K(\boldsymbol{Z},\boldsymbol{Z})+C]^{-1}K(\boldsymbol{Z},\boldsymbol{Z^{*}})\nonumber,
\end{align}
    where $\boldsymbol{Z^*}=$\{$z_i^*$\} represents the vector of redshifts to be predicted, and $\boldsymbol{y}=\{y_i\}$ denotes the vector of observations. The covariance matrix of the observed data is denoted by $C$, where $C=\text{diag}\{\sigma_i^2\}$ and $\sigma_i$ represents the error value corresponding to each data point. In addition, the calculation of the covariance matrix depends on the sampling of the hyperparameters ($\sigma_f^2$,$l$). To prevent the program from encountering a ``singular matrix" error, Cholesky decomposition is used instead of solving the inverse of the matrix directly throughout the Gaussian process.

The final reconstruction of the GP can be influenced by any variables that impact the computation of the covariance matrix. This is because the covariance matrix $\emph{K\textbf{(Z,Z)}}$ defines both the mean function and covariance function of the GP. The shape of the kernel function can directly affect the outcome. In cosmology research, it is commonly assumed that the kernel function should only depend on the distance between different locations. Therefore, the RBF kernel is often used due to its simplicity:
\begin{itemize}
    \item {Radial Basis Function (RBF)}
    \begin{equation}\label{RBF}
    k(z_i,z_j)=\sigma_f^2\exp({-\frac{(z_i-z_j)^2}{2l^2}}).
\end{equation}
\end{itemize}

In Equation \eqref{RBF}, the hyperparameters $\sigma_f^2$ and $l$ are typically used in kernel functions. However, there are other kernels besides the RBF kernel that can be used in practical research, such as the Cauchy kernel, the Matérn($\nu$=5/2) kernel, the Matérn($\nu$=7/2) kernel, and the Matérn($\nu$=9/2) kernel. These kernels will also be investigated in this study. Here is a list of the additional kernels:
\begin{itemize}
    \item {Cauchy kernel (CHY)}
    \begin{align}\label{CHY}
        k(z_i,z_j)=\sigma_f^2(1+\frac{(z_i-z_j)^2}{2l^2})^{-1},
    \end{align}
    \item {Matérn($\nu$=5/2) kernel (M52)}
    \begin{align}\label{M52}
        k(z_i,z_j)&=
        \sigma_f^2\exp({-\sqrt{5}\frac{|z_i-z_j|}{l}})\nonumber \\
        &(1+\sqrt{5}\frac{|z_i-z_j|}{l}+5\frac{(z_i-z_j)^2}{3l^2}),
    \end{align}
    \item {Matérn($\nu$=7/2) kernel (M72)}
    \begin{align}\label{M72}
        &k(z_i,z_j)=
        \sigma_f^2\exp({-\sqrt{7}\frac{|z_i-z_j|}{l}})(1+\nonumber \\
        &\sqrt{7}\frac{|z_i-z_j|}{l}+14\frac{(z_i-z_j)^2}{5l^2}+7\sqrt{7}\frac{|z_i-z_j|^3}{15l^3}),
    \end{align}
    \item {Matérn($\nu$=9/2) kernel (M92)}
    \begin{align}\label{M92}
        &k(z_i,z_j)=\sigma_f^2\exp({-3\frac{|z_i-z_j|}{l}})
        (1+3\frac{|z_i-z_j|}{l}\nonumber \\
        &+27\frac{(z_i-z_j)^2}{7l^2}+18\frac{|z_i-z_j|^3}{7l^3}+27\frac{(z_i-z_j)^4}{35l^4}).
    \end{align}
\end{itemize}

The kernels mentioned earlier are commonly used in various research fields. However, the M92 kernel is of particular interest as studies have revealed it to be the most effective kernel for constraining the equation of state parameter of dark energy \citep{seikel2013optimising}, and it is regularly employed in dark energy surveys. Thus, we have also incorporated the M92 kernel in our study.

The impact of the hyperparameters ($\sigma_f^2$ and $l$, as explained earlier) on the computation of the covariance matrix should not be overlooked. These hyperparameters must be trained for a complete GP in order to determine their optimal values. In cosmological research, the Log Marginal Likelihood (LML) is a widely used method for hyperparameter training. The goal of hyperparameter training is to find the combination of hyperparameters that correspond to the maximum LML, which is then used as the final set of hyperparameters to be substituted into the GP for the final result. The mathematical form of LML is as follows:

\begin{align}\label{LML}
     \ln\mathcal{L}&=-\frac{1}{2}\boldsymbol{y}^{\top}[K(\boldsymbol{Z},\boldsymbol{Z})+C]^{-1}\boldsymbol{y}\nonumber \\
     &-\frac{1}{2}\ln|K(\boldsymbol{Z},\boldsymbol{Z})+C|-\frac{n}{2}\ln 2\pi,
\end{align}
where $n$ is the dimension of the vector $\boldsymbol{Z}$. It's important to note that other methods can also be used to obtain hyperparameters. When the LML reaches its maximum value, the relevant hyperparameters yield the most likely manifestation of the function. In practice, almost all GPs are carried out by maximizing the LML function.

To illustrate how our data can be used to complete the GP, let's consider an example. Suppose we have an $H(z)-z$ dataset with $n$ data points, and we assume that the hyperparameters ($\sigma_f^2$ and $l$) are sampled from a given distribution. With the kernel function chosen, we can construct a kernel function matrix $K(\boldsymbol{X,X})$ for the $n$ redshift values contained in the data, where $\boldsymbol{X}$ is a column vector consisting of the redshifts of the observation points. We also have $n'$ prediction points at which we want the GP to predict the value of the Hubble parameter $H(z)$, and the vector consisting of these $n'$ redshifts is denoted as $\boldsymbol{X^{*}}$. By combining the error $\sigma_n$ at each data point and the actual observed Hubble parameter $H(z)$ obtained (here we use $\boldsymbol{y}$ to refer to the column vector made up of observations), we can start the GP:
\begin{align}
    L&=\text{cholesky}(K(X,X)+\sigma_n^2I), \nonumber\\
    \boldsymbol{\alpha}&=L^{\top}\backslash(L\backslash\boldsymbol{y}),
\end{align}
where $I$ is the identity matrix. It has the same dimension as $K(\boldsymbol{X,X})$. Now we can get the mean values:
\begin{equation}
    \overline{\boldsymbol{f^{*}}}=K(\boldsymbol{X^{*},X})\boldsymbol{\alpha}.
\end{equation}

Next we construct the variance:
\begin{align}
    \boldsymbol{v}&=L\backslash K(\boldsymbol{X^{*},X}), \nonumber\\
    \mathrm{cov}(\boldsymbol{f^{*}})&=K(\boldsymbol{X^*,X^*})-\boldsymbol{v}^{\top}\boldsymbol{v}.
\end{align}
Also, we can get the log-marginal-likelihood:
\begin{equation}
    \ln\mathcal{L}=-\frac{1}{2}\boldsymbol{y}^{\top}\boldsymbol{\alpha}-\frac{1}{2}\ln|K(\boldsymbol{X},\boldsymbol{X})+\sigma_n^2I|-\frac{n}{2}\ln 2\pi.
\end{equation}
It's important to note that in the previous discussion of the GP principle, we replace $\sigma_n^2I$ with $C$. However, this is only a convenient substitution for presentation purposes. If you need to write an actual GP algorithm, the steps mentioned above are what you need to follow. This is also the step where we apply the principles of GP to model the data \citep{williams2006gaussian}.

In this study, we evaluate the effectiveness of various GP kernels using three datasets that correspond to three different cosmological probes: cosmic chronometer (CC), type Ia supernovae (SNIa), and Gamma Ray Burst (GRB). The CC dataset is presented in Table \ref{tab:H(z)}, while the SNIa data, obtained from \cite{scolnic2018complete}, are binned and consist of 40 data points. The original dataset contains 1048 points, but the binned version is used in this study. The covariance matrix, provided by \cite{scolnic2018complete}, is used directly in the GP calculations. The GRB dataset, obtained from \cite{demianski2017cosmology}, has a large number of data points in the high-redshift zone, making it an excellent source of information about the high-redshift universe. We choose not to transform the GRB and SNIa data into redshift with other parameters since this process is model-dependent \citep{yang2013improved}. We aim to minimize the reliance on a specific cosmological model. Furthermore, the uncertainty in the luminosity distance of GRB data necessitates the use of a specific likelihood function \citep{reichart2001possible}, which is not pertinent to this study.

\begin{figure*}[htbp]
    \centering
    \includegraphics[scale=0.4]{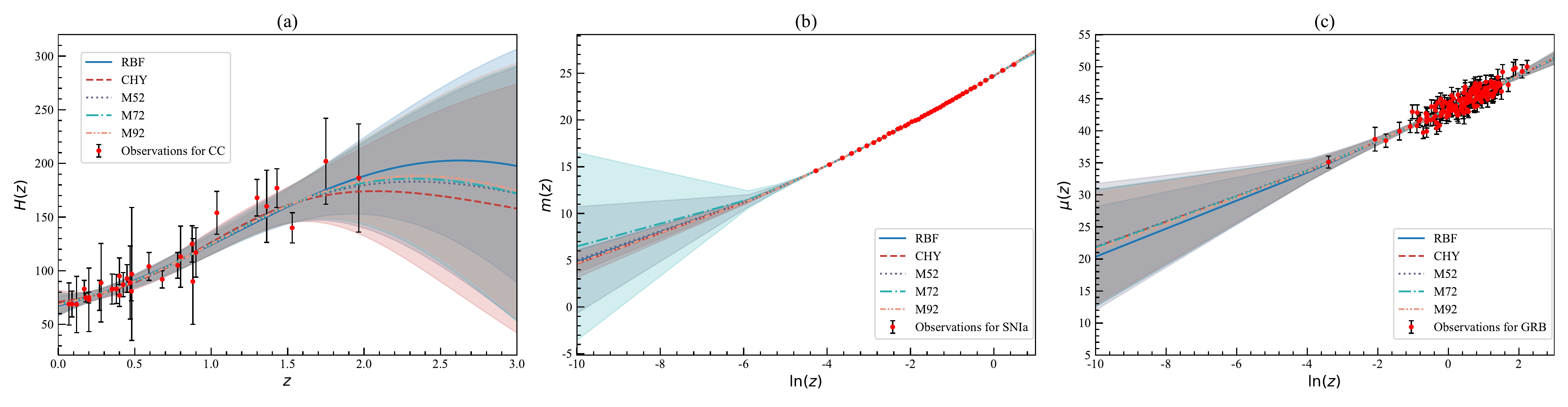}
    \caption{Reconstructed images of GP obtained using scikit-learn. For all 3 graphs, the different coloured curves represent the curves of the different kernel reconstructions, the red data points with error bars represent the observed data and the shaded regions corresponding to the colour of the curves indicate the range of 95\% confidence intervals of the reconstruction. For each of them, we have the following descriptions.\\ (a): Reconstruction for CC. The horizontal axis $z$ represents the redshift and the vertical axis $H(z)$ represents the value of the Hubble parameter obtained from the observation.  \\(b): Reconstruction for SNIa(binned). The horizontal axis $\ln(z)$ represents the redshift value after logarithmization and the vertical axis $m(z)$ represents the distance modulus of the supernovae. \\(c): Plot for reconstruction of GP using GRB dataset. The horizontal axis has the same meaning as in Fig.(b). The vertical axis $\mu(z)$ represents the distance modulus of the GRB event.}
    \label{gpcc}
\end{figure*}

We use the \verb|scikit-learn| \footnote{\url{https://scikit-learn.org/stable/index.html}} module to demonstrate the general GP reconstruction generated using LML training hyperparameters. The hyperparameters are optimized by LML, where the constant-value and length-scale correspond to $\sigma_f^2$ and $l$, respectively. Figure \ref{gpcc} shows that different kernel selections result in distinct curves after reconstruction, but it is challenging to infer which performs better from the graphs. To address this issue, we propose using the Approximate Bayesian Computation (ABC) Rejection technique. The results of the GP reconstruction of the observed data will also be needed when using ABC Rejection, where the GP will be additionally programmed instead of using the one in \verb|scikit-learn|.
\begin{deluxetable}{cccc}
\caption{32 $H(z)$ Measurements from the Cosmic Chronometer Method}\label{tab:H(z)}
\tablewidth{0pt}
\tablehead{
\colhead{$z$} & \colhead{$H(z)$} & \colhead{$\sigma_H$} &
\colhead{References}
}
\startdata
0.09 & 69 & 12 &\cite{Jimenez_2003}\\
\hline
0.17 & 83 & 8 & \\
0.27 & 77 & 14 & \\
0.4 & 95 &17 & \\
0.9 & 117 & 23 & \\
1.3 & 168 & 17 & \cite{PhysRevD.71.123001}\\
1.43 & 177 & 18 & \\
1.53 & 140 & 14 & \\
1.75 & 202 & 40 & \\
\hline
0.48 & 97 & 62 & \cite{Stern_2010}\\
0.88 & 90 & 40 & \\
\hline
0.1791 & 75 & 4& \\
0.1993 & 75 & 5 & \\
0.3519 & 83 & 14 & \\
0.5929 & 104 & 13 & \\
0.6797 & 92 & 8 & \cite{Moresco_2012}\\
0.7812 & 105 & 12 & \\
0.8754 & 125 & 17 &  \\
1.037& 154 & 20 & \\
\hline
0.07 & 69 & 19.6 & \\
0.12 & 68.6 & 26.2 & \cite{Zhang_2014}\\
0.2 & 72.9 & 29.6 & \\
0.28 & 88.8 & 36.6 & \\
\hline
1.363 & 160 & 33.6 & \cite{10.1093/mnrasl/slv037}\\
1.965 & 186.5 & 50.4 & \\
\hline
0.3802 & 83 & 13.5 & \\
0.4004 & 77 & 10.2 & \\
0.4247 & 87.1 & 11.2 & \cite{Moresco_2016}\\
0.4497 & 92.8 & 12.9 & \\
0.4783 & 80.9 & 9& \\
\hline
0.47 & 89 & 34 & \cite{10.1093/mnras/stx301}\\
\hline
0.80 & 113.1 & 28.5 & \cite{Jiao2022new}\\
\enddata
\tablecomments{The observational Hubble parameter $H(z)$ measurements of CC (in units of km$\cdot$s$^{-1}\cdot$Mpc$^{-1}$) and their errors $\sigma_H$ at redshift $z$. Here we have 32 data points in total.}
\end{deluxetable}

\subsection{Approximate Bayesian Computation(ABC) Rejection}\label{ABC}
The foundation of the standard Bayesian inference procedure lies in the prior distribution of the parameters and the likelihood function, which use the Bayes rule to determine the posterior distribution. The Bayes rule is expressed as follows:
\begin{equation}
    P(\theta|\boldsymbol{y})=\frac{P(\theta)P(\boldsymbol{y}|\theta)}{\int P(\boldsymbol{y}|\theta)P(\theta)\mathrm{d}\theta},
\end{equation}
where $P(\theta|\boldsymbol{y})$ denotes the posterior distribution, which is obtained through Bayesian analysis. Given the observed dataset $\boldsymbol{y}$, it represents the probability distribution of the parameter $\theta$ value. The researcher can determine the prior distribution of the parameter $\theta$ denoted by the prior $P(\theta)$. The prior is the probability function, or $P(\boldsymbol{y}|\theta)$. The normalizing constant is $\int P(\boldsymbol{y}|\theta)P(\theta)\mathrm{d}\theta$, which is referred to as ``evidence". For simplicity, we leave it at 1. Bayes rule is currently expressed as:
\begin{equation}\label{Bayes}
    P(\theta|\boldsymbol{y})=P(\theta)P(\boldsymbol{y}|\theta).
\end{equation}

Bayesian analysis, based on Equation \eqref{Bayes}, when combined with the MCMC approach, has been widely used in cosmological parameter estimation (CPE) and has become a standard way of analysis in CPE\citep{zuntz2015cosmosis}. Bayesian analysis can also be used for model selection\citep{trotta2008bayes}. Furthermore, in our work, choosing kernels can be seen as a form of model selection, and thus, our conception of model selection is also founded on Bayesian analysis. When applying Bayesian analysis to model selection, Equation \eqref{Bayes} must be modified to take the following form:
\begin{equation}\label{Model Bayes}
    P(M_i|\boldsymbol{y})=P(M_i)P(\boldsymbol{y}|M_i).
\end{equation}

The posterior distribution that we require represents the probability of the collection of observations being produced by the model $M_i$. However, directly applying Equation \eqref{Model Bayes} is difficult since the likelihood function $P(\boldsymbol{y}|M_i)$ cannot be represented analytically when the various kernels are viewed as distinct models.

It is important to note that although GP assumes that the reconstructed functions in the model follow a Gaussian likelihood distribution, this does not imply that $P(\boldsymbol{y}|M_i)$ is a Gaussian likelihood. Here $M_i$ represents different kernels. In fact, GP assumes that the vector of observations is obtained from a Gaussian distribution in the function space\citep{seikel2012reconstruction}, and that the mean of this Gaussian distribution is our prediction $\overline{\boldsymbol{f^{*}}}$. Hence, if we use Bayes' theorem for kernel selection, the resolved form of its likelihood function remains unknown to us. This is also mentioned in \cite{bernardo2021towards}. Since we cannot represent the likelihood function analytically, the traditional MCMC approach is not applicable to our problem.

As an alternative, we choose Approximate Bayesian Computation (ABC) Rejection, which has the benefit of not requiring the specification of a likelihood function\citep{turner2012tutorial}. This quality makes the ABC Rejection approach popular in cosmology (\cite{jennings2017astroabc};\cite{weyant2013likelihood};\cite{ishida2015cosmoabc}). In ABC Rejection, the likelihood function is approximated by using frequencies to estimate probabilities, allowing us to derive the posterior distribution. Here, we sample the model's parameters multiple times, referring to each sample as a particle. We then establish suitable screening criteria and calculate the fraction of particles that pass the screening relative to the total number of samples to derive the frequency and, consequently, the probability. To apply the ABC Rejection, we treat the kernel function as a model $M$, and sample the hyperparameters $\sigma_f^2$ and $l$ as parameters in the model $M$. The following are the steps to use the ABC Rejection method\citep{toni2009tutorial}:

\begin{itemize}
    \item To ensure uniformity across all kernels studied, we assign the same prior probability, $p(M_i)$, to each kernel. From this collection of kernels, we select a specific kernel, $k(z_i,z_j)$, as our model $M$. The hyperparameters, $(\sigma_f^2,l)$, which are treated as a 'particle', are then sampled from the prior distribution.
    
    \item Using the sampled hyperparameters $(\sigma_f^2,l)$ and the selected kernel $k(z_i,z_j)$, we obtain the mean of the reconstructed function, denoted by $\overline{\boldsymbol{f^*}}$. This reconstructed function is generated using GP. It is important to note that the mean of the reconstructed function $\overline{\boldsymbol{f^*}}$ is equivalent to a simulated dataset.
    
    \item We then calculate the distance, denoted by $d$, between the simulated dataset $\overline{\boldsymbol{f^*}}$ and the observed dataset $\boldsymbol{y}$ by substituting both into the distance function $d(x,y)$.
    
    \item To determine whether the sampled hyperparameters $(\sigma_f^2,l)$ are acceptable, we set a threshold, denoted by $\varepsilon$, and compare the calculated distance $d$ with $\varepsilon$. If $d\leq\varepsilon$, we accept $(\sigma_f^2,l)$ and add it to the count. Otherwise, we discard $(\sigma_f^2,l)$ and do not count it.
    
    \item By repeating the above steps, we can obtain the posterior distribution $p(\boldsymbol{y}|M_i)$ that we want to calculate. Specifically, we sample $(\sigma_f^2,l)$ multiple times from a particular distribution and count the number of particles that satisfy $d\leq\varepsilon$. Finally, we solve for the ratio of that number to the total number of samples. When $\varepsilon$ is sufficiently small to indicate obvious differences, this ratio corresponds to the posterior distribution $p(\boldsymbol{y}|M_i)$. 
\end{itemize}

As can be observed from the above stages, when computing the posterior distribution using ABC Rejection for the results, the selection of the distance function $d(x,y)$, the distribution of the hyperparameters sampled, and the sampling technique chosen are all critical. In our research, we utilized hyperparameters from a uniform distribution, which is available in the \verb"abcpmc" package\citep{akeret2015approximate}.

In practice, many previous studies combined importance sampling algorithms to assign weights to the particles, gradually reducing the threshold to conserve computational resources for the ABC Rejection algorithm. The SMC (Sequential Monte Carlo) algorithm\citep{toni2010simulation} and the PMC (Population Monte Carlo) algorithm\citep{cappe2004population} are two typical algorithms that are computed iteratively, and the threshold is reduced after each generation according to the weight of the particles that have not yet been filtered out. Therefore, adaptive algorithms for weight calculation can also be applied(\cite{beaumont2009adaptive};\cite{bonassi2015sequential}). Although the ABC-PMC algorithm is usually employed for parameter inference without likelihood functions\citep{kilbinger2012cosmopmc} and less frequently for model selection, the ABC-SMC algorithm is very effective when dealing with large-scale data and has also been used for model selection(\cite{abdessalem2017automatic};\cite{bernardo2021towards}).

To assign weights to the particles, several previous studies integrated importance sampling approaches, gradually lowering the threshold to conserve computational resources for the ABC Rejection procedure. The SMC (Sequential Monte Carlo) algorithm\citep{toni2010simulation} and the PMC (Population Monte Carlo) algorithm\citep{cappe2004population} are two examples of these algorithms. They are calculated iteratively, and the threshold is lowered in accordance with the weight of the particles that haven't yet been filtered out after each generation. As a result, adaptive algorithms can also be used to calculate weights (see \cite{beaumont2009adaptive};\cite{bonassi2015sequential}). While the ABC-PMC algorithm is typically employed for parameter inference without likelihood functions\citep{kilbinger2012cosmopmc} and less frequently for model selection, the ABC-SMC method has been used for model selection and is particularly efficient when working with large datasets. However, since the dataset we use is not particularly large, and the ABC-SMC technique only returns "which is the best kernel" and cannot compare the magnitude of each kernel relative to the data, we do not intend to employ it in our study. In fact, even without importance sampling in our investigation, we can still obtain consistent results. We only need to consider how different distance functions affect the posterior distribution. Section \ref{dis_f} describes the distance functions used in the study.

\subsection{Distance Function} \label{dis_f}

The selection of an appropriate distance function is critical for Approximate Bayesian Computation (ABC) analysis as different functions may indicate varying levels of statistical significance between simulated and real datasets. Previous research in this field has typically used the likelihood function (LML) as the distance function (\cite{abdessalem2017automatic}; \cite{bernardo2021towards}). LML is frequently used to evaluate the impact of hyperparameter values on the fit, making it a reasonable choice as one of the distance functions. However, selecting hyperparameter values is not limited to LML optimization and other methods can also be used \citep{gomez2018h0}. In our work, we selected several distance functions, including LML, $\chi^2$ estimation, and bias, to compare their performance in ABC analysis:

\begin{itemize}
    \item {$\chi^2$\citep{bernardo2021towards}}
    \begin{align}\label{chi2}
        \chi^2=d(\overline{\boldsymbol{f^*}},\boldsymbol{y})=(\overline{\boldsymbol{f^*}}-\boldsymbol{y})^T\Sigma_y^{-1}(\overline{\boldsymbol{f^*}}-\boldsymbol{y}),
    \end{align}
    \item {Bias}
    \begin{align}\label{bias}
        d(\overline{f^*},y)=\frac{\sum_{i=1}^n|y_i-\overline{f^*}_i|}{n}.
    \end{align}
\end{itemize}

Optimizing the residuals of the data is equivalent to minimizing Equation \eqref{chi2}. This equation considers the need to minimize the sum of squares of the residuals while taking into account the weighting of the inverse error. Based on the residuals, Equation \eqref{chi2} provides a criterion for assessing the model's merit, where a smaller value of $\chi^2$ indicates better agreement between the simulated and observed data.

Equation \eqref{bias} provides the average of Euclidean distances between the simulated and observed datasets and represents the expectation of the difference between the projected and actual values of the model, also known as bias. The value of bias indicates how well the model fits the data, and the lower its value, the closer the mean value of the simulated data is to the observed data.

By combining these three distance functions, we provide three different methods for filtering particles, and the ABC Rejection results obtained from these methods offer a more comprehensive answer to the question of which kernel performs best in ABC analysis.

\subsection{Bayes Factor}\label{Bayesfactor}
The Bayes factor, denoted by $K$, is a measure of the support for one statistical model over another, obtained by taking the ratio of their likelihoods\citep{MOREY20166}. In our work, we use the Bayes factor to assess the dependence of different datasets on the kernel. Unlike traditional hypothesis testing, which only allows for acceptance or rejection of a hypothesis, the Bayes factor evaluates the evidence in support of a hypothesis. Hence, the Bayes factor not only identifies the best model among competing kernels but also quantifies how much better it is than the others. For two alternative models $M_1$ and $M_2$ with observed data $\boldsymbol{y}$, the plausibility of the models is evaluated by the Bayes factor $K$, which is given by Equation \eqref{Bayes_factor}:
\begin{equation}\label{Bayes_factor}
    K_{12}=\frac{P(\boldsymbol{y}|M_1)}{P(\boldsymbol{y}|M_2)}=\frac{P(M_1|\boldsymbol{y})}{P(M_2|\boldsymbol{y})}\frac{P(M_2)}{P(M_1)}.
\end{equation}

In fact, the Bayes factor is the ratio of the evidence from the two models. But in order to be able to use the results of ABC Rejection, we have transformed the original form of the Bayes factor into Equation \eqref{Bayes_factor} to use the results of ABC Rejection (see Section \ref{nestsample} for a direct calculation of the evidence). When the prior probabilities of the two models are equal ($P(M_1)=P(M_2)$), the Bayes factor $K$ represents the ratio of the posterior probability of $M_1$ and $M_2$. We compute the same prior probability for the two kernels when we calculate the Bayes factor between them. The procedure only takes into account the ratio of the posterior distributions of the two kernels.

In Equation \eqref{Bayes_factor}, a value of $K>1$ implies that $M_1$ is more supported by the data than $M_2$. The scale of $K$ has a quantitative interpretation based on probability theory\citep{jeffreys1998theory}, as shown in Table \ref{tab:K}. We use this table to determine the strength of evidence for each of the two kernels.

\begin{deluxetable}{ccc}
\caption{Correspondence between the Value of $K$ and the Strength of the Evidence}\label{tab:K}
\tablewidth{0pt}
\tablehead{
\colhead{$K_{12}$} & \colhead{$\log_{10}K$} & \colhead{Strength of Evidence}
}
\startdata
$<10^0$ & $<$0 & 	Negative (supports $M_2$)\\
$10^0$ to $10^{1/2}(\approx3.162)$ & 0 to 1/2 & Barely worth mentioning \\
$10^{1/2}(\approx3.162)$ to $10^1$ & 1/2 to 1 & Substantial\\
$10^1$ to $10^{3/2}(\approx31.623)$ & 1 to 3/2 & Strong\\
$10^{3/2}(\approx31.623)$ to $10^2$ & 3/2 to 2 & Very strong\\
$>10^2$ & $>$2 & Decisive\\
\enddata
\tablecomments{$K$ denotes the value of the Bayes factor and $\log_{10}K$ denotes the result of taking the usual logarithm of $K$. The third column indicates the strength of the evidence in the corresponding range of $K$ values. Subsequent analyses of the Bayes factor results are based on this table. This table gives the criteria for quantitatively evaluating the relative reliability between two models.}
\end{deluxetable}

\subsection{Nested Sampling}\label{nestsample}
Nested sampling is a method originally proposed by John Skilling to approximate the evidence in Bayes' theorem\citep{10.1214/06-BA127}. Bayes' theorem takes the form of:
\begin{align}\label{evidence}
    P(\theta|D,M)&=\frac{P(D|\theta,M)P(\theta|M)}{P(D|M)},\nonumber\\
    P(D|M)&=\mathcal{Z}=\int P(D|\theta,M)P(\theta|M)\mathrm{d}\theta.
\end{align}
In Bayes' theorem, $P(D|M)$ is referred to as the ``evidence" and is an unresolvable representation of the likelihood function, as mentioned in ABC Rejection. $D$ represents our observed data, while $\theta$ represents the parameter of the model $M$ (which may include additional parameters). To make model comparisons, the easiest approach is to compare the size of the evidence corresponding to different models with the same prior probability. Nested sampling is the method by which we can obtain the evidence directly.

To perform model comparisons using nested sampling, we first rewrite Bayes' theorem as follows\citep{handley2015polychord}:
\begin{align}\label{modelcom}
    P(M_i|D)&=\frac{P(D|M_i)P(M_i)}{P(D)},\nonumber\\
    &=\frac{\mathcal{Z}_i\pi_i}{\Sigma_j\mathcal{Z}_j\pi_j}.
\end{align}
In Equation \eqref{modelcom}, $\pi$ refers to the priors of the model $M_i$. In practice, it is often assumed that $\pi$ is equal for each model to be compared, and so Equation \eqref{modelcom} is not necessary for model comparison. The only integral that needs to be calculated is that expressed in Equation \eqref{evidence}, and the model with the greater value of evidence is deemed to be superior in the comparison.

The principle of nested sampling can be briefly described as follows\citep{handley2015polychord}. We begin by sampling $n_{\text{live}}$ points from the prior distribution. We then define a prior volume $X(\mathcal{L})$:
\begin{equation}
    X(\mathcal{L})=\int_{\mathcal{L}(\theta)>\mathcal{L}}\pi(\theta)\mathrm{d}\theta,
\end{equation}
where $\mathcal{L}$ is a likelihood we set. At each iteration $i$, we remove the points with likelihood lower than $\mathcal{L}$ and replace them with new live points. According to Equation \eqref{evidence}, the integral now takes the form of:
\begin{equation}
    \mathcal{Z}=\int_0^1\mathcal{L}(X)\mathrm{d}X.
\end{equation}
The integral can be written in the form of a summation to the point to get the approximation of the evidence. More details are shown in \cite{10.1214/06-BA127}, \cite{handley2015polychord} and \cite{handley2015polychord_2}.

To implement nested sampling, we can use the Python package \verb"Dynesty"\citep{speagle2020dynesty}, which is easy to install and use. With \verb"Dynesty", we input a prior and a log-marginal-likelihood, and its internal nested sampler is called to complete the sampling. For GP, the prior used is not the same as the prior of hyperparameters; it refers to the reconstructed data $\boldsymbol{y}$. However, the log-marginal-likelihood is the same. In \verb"Dynesty", the log-marginal-likelihood takes the form of $P(D|\theta,M)$, and for GP, the LML is $P(\boldsymbol{y}|\sigma_f,l,\boldsymbol{X})$. Here, $\boldsymbol{X}$ is equal to $\theta$, while $\sigma_f$ and $l$ describe the model $M$. Thus, we can directly use the LML in GP as the likelihood in \verb"Dynesty".

Another important option in \verb|Dynesty| is the choice of a specific bounding. We choose ``Multiple ellipsoid bounds", as it is the most frequently used method\citep{feroz2009multinest}.

In \verb"Dynesty", the program output for evidence is usually $\ln\mathcal{Z}$. As the essence of the Bayes factor is the division of the evidence from two models, $\ln\mathcal{Z}$ becomes a subtraction between different models' $\ln\mathcal{Z}$:
\begin{equation}\label{lnz}
    \ln K_{12}=\ln P(\boldsymbol{y}|M_1)-\ln P(\boldsymbol{y}|M_2)
    =\ln\mathcal{Z}_1-\ln\mathcal{Z}_2.
\end{equation}
As in Table \ref{tab:K}, such a comparison table will be given for $\ln K_{12}$. The relationship between this difference and the strength of model support is shown in the Tabel \ref{tab:lnK}\citep{sarro2012astrostatistics}.

\begin{deluxetable}{ccc}
\caption{Correspondence between the Value of $\ln K_{12}$ and the Strength of the Evidence}\label{tab:lnK}
\tablewidth{0pt}
\tablehead{
\colhead{$|\ln K_{12}|$} & \colhead{odds} & \colhead{Strength of evidence}
}
\startdata
$<1.0$ & $\lesssim 3:1$ & Inconclusive \\
$1.0$ & $\sim 3:1$ & Weak evidence \\
$2.5$ & $\sim 12:1$ & Moderate evidence\\
$5.0$ & $\sim 150:1$ & Strong evidence\\
\enddata
\end{deluxetable}

\section{Results} \label{sec:results}
\subsection{ABC Rejection with Different Distance Functions}

\begin{figure*}[htbp]
\begin{center}
\includegraphics[scale=0.4]{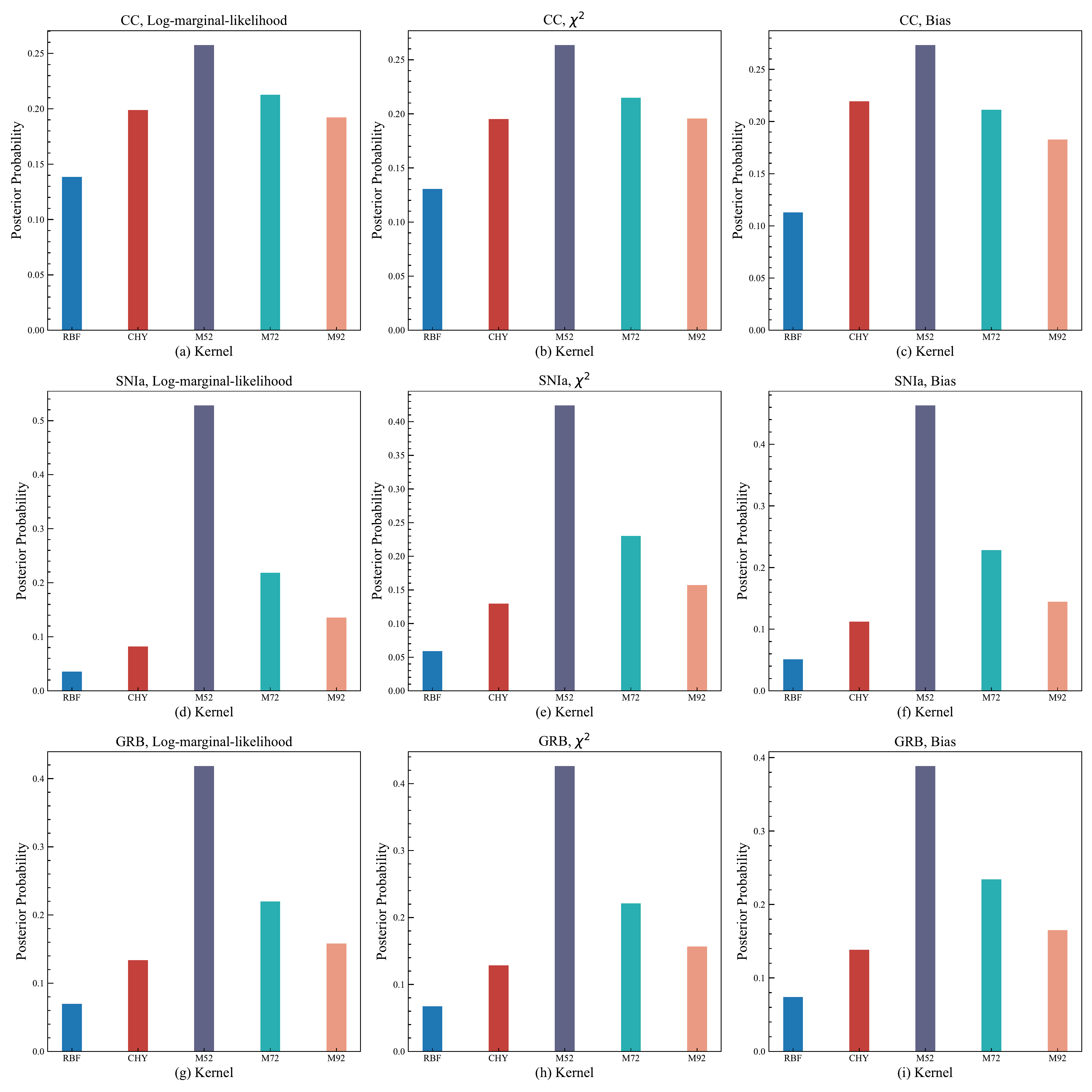}
\end{center}
\caption{For each subplot, we utilize a bar plot format to display the posterior distribution of the five kernels. The horizontal axis for each subplot denotes the type of kernel, while the vertical axis indicates the likelihood that the observed data is created by them. Normalization is applied to the probability.}
\label{ABCRejection}
\end{figure*}

In this investigation, we employ three different types of data, apply the ABC rejection method to each type of data, and use three different distance functions in the computations, resulting in nine graphs presented in Figure \ref{ABCRejection}.

It is important to note that the size of the threshold is not chosen arbitrarily. Setting the threshold too high would obscure the differences between specific kernel functions, while setting the threshold too low would result in not only a small number of particles in each kernel function, but also particles that are very close to one another. For instance, when the threshold is set too low, the RBF kernel produces only two particles that are closer to the observed data than the threshold distance, while the M52 kernel produces four particles. Although the posterior distribution of the M52 kernel is twice as large as that of the RBF kernel in terms of normalized probability, the difference between the two is not significant. Therefore, an incorrect judgment would be made if the threshold is set too low.

To address this issue, we continuously adjust the threshold until we reach the final result. When the posterior distributions of the individual kernels undergo significant changes when the threshold is set to $\varepsilon$, but do not differ significantly when the threshold is greater than $\varepsilon$, we consider $\varepsilon$ to be the appropriate threshold. When the previously observed differences are preserved when the threshold is set to a value less than $\varepsilon$, we consider $\varepsilon$ to be the correct threshold. It is worth mentioning that there are no circumstances where the differences in the posterior distributions of the kernels change when the threshold is decreased further.

After resolving the threshold selection problem, our next step is to reduce as much randomness as possible from the results. We ensure that the sampling opportunities are equal for all 5 kernels, meaning that their prior probabilities are all 20\%. This is reflected in the sampling of $2\times10^5$ samples for each kernel in each scenario. At the end of every $10^4$ samples, we record each kernel's posterior distribution, resulting in 100 posterior probabilities for each kernel in each example. The posterior distribution for each kernel function in Figure \ref{ABCRejection} is obtained by averaging these 100 posterior probabilities.

As shown in Figure \ref{ABCRejection}, the M52 kernel has the largest posterior distribution in all situations, indicating that the GP reconstructed by the M52 kernel is the closest to the real data and is therefore the best model. The extent to which M52 outperforms the other kernels varies between subplots.

It is worth noting that the RBF kernel, which is commonly used by astronomers, has a larger bias in all reconstruction instances than the other kernels. On the other hand, the M52 kernel reconstruction fits the data much better than the RBF kernel.

Based on the ABC Rejection results, we can conclude that M52 is the best kernel for the purpose of this study. Similar results are presented in \cite{bernardo2021towards}, but there are no results for M52 compared to other models. The strength of the evidence is discussed in Section \ref{Bayesresults} in comparison to other kernels.

\subsection{Bayes Factor $K$ Between Different Kernels (in ABC Rejection)}\label{Bayesresults}

We generate heat maps to visualize the magnitude of the Bayes factor $K$ between the kernels for each of the nine cases depicted in Figure \ref{ABCRejection}. A colorbar is used to represent the Bayes factor values, with significant values of 1, 3.16, and 10 labeled.

\begin{figure*}[htbp]
\begin{center}
\includegraphics[scale=0.4]{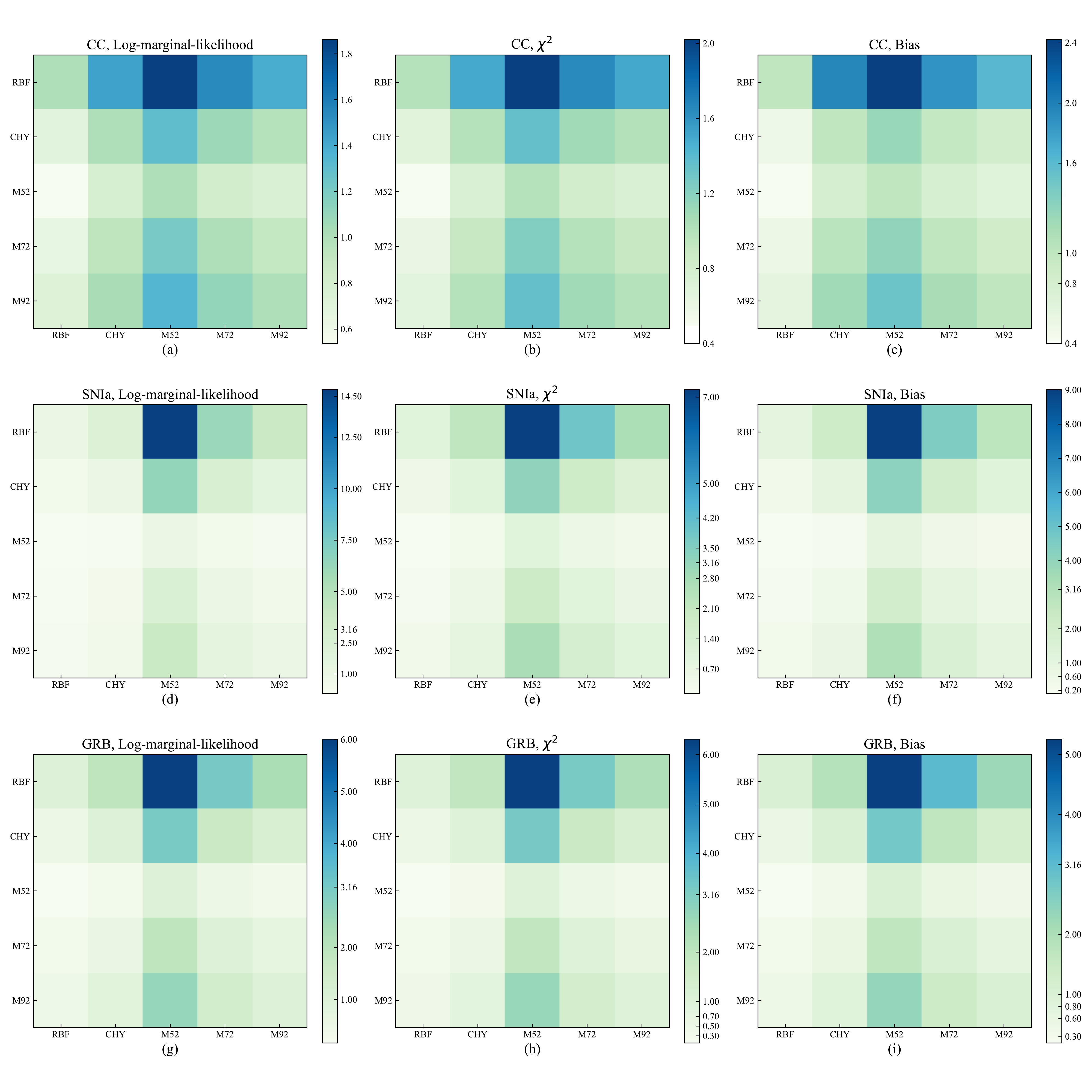}
\end{center}
\caption{The Bayes factor between every 2 kernels is visualized using this heat map to represent the strength of evidence level of the observed data between any two kernels. As an example, the value corresponding to the color block in the second column of the first row on the colorbar indicates the Bayes factor of CHY with respect to RBF. That is, $P(M_{\text{CHY}}|\boldsymbol{y})/P(M_{\text{RBF}}|\boldsymbol{y})$(According to Equation \eqref{Bayes_factor}). The dataset and distance function related to the posterior distribution utilized for the Bayes factor computation for that subplot are indicated in the subplot's title.}
\label{Bayefac}
\end{figure*}

Figure \ref{Bayefac} shows that the M52 kernel significantly outperforms the RBF kernel, while the difference in performance between M52 and the other kernels is not significant, except for the comparison with RBF.

For the CC dataset, the strength of evidence for M52 is only at the ``Barely worth mentioning" level compared to RBF and other kernels. However, the Bayes factor of M52 for RBF is the highest among all cases.

When using the SNIa dataset, the evidence strength of M52 is at the ``Strong" level compared to RBF, occasionally reaching ``Strong" when compared to other kernels but mostly staying at ``Substantial" and ``Barely worth mentioning". The M72 kernel is the second-best performing kernel after M52.

For the GRB dataset, the evidence strength for M52 relative to RBF is ``Substantial" with LML, Bias, and $\chi^2$ distance functions. Other kernels have evidence ranging from ``barely worth mentioning" to ``strong", with M72 being the second-best performing kernel after M52.

Although the M52 kernel outperforms all other kernels, the strength of the evidence is not always decisive. Therefore, in most cases, replacing M52 with a different kernel would not affect the GP reconstruction significantly. However, when choosing between M52 and RBF, we prefer to use M52 as the commonly used kernel in GP.

\subsection{Nested Sampling: Evidence and Bayes factors}

\begin{figure*}[htbp]
\begin{center}
\includegraphics[scale=0.65]{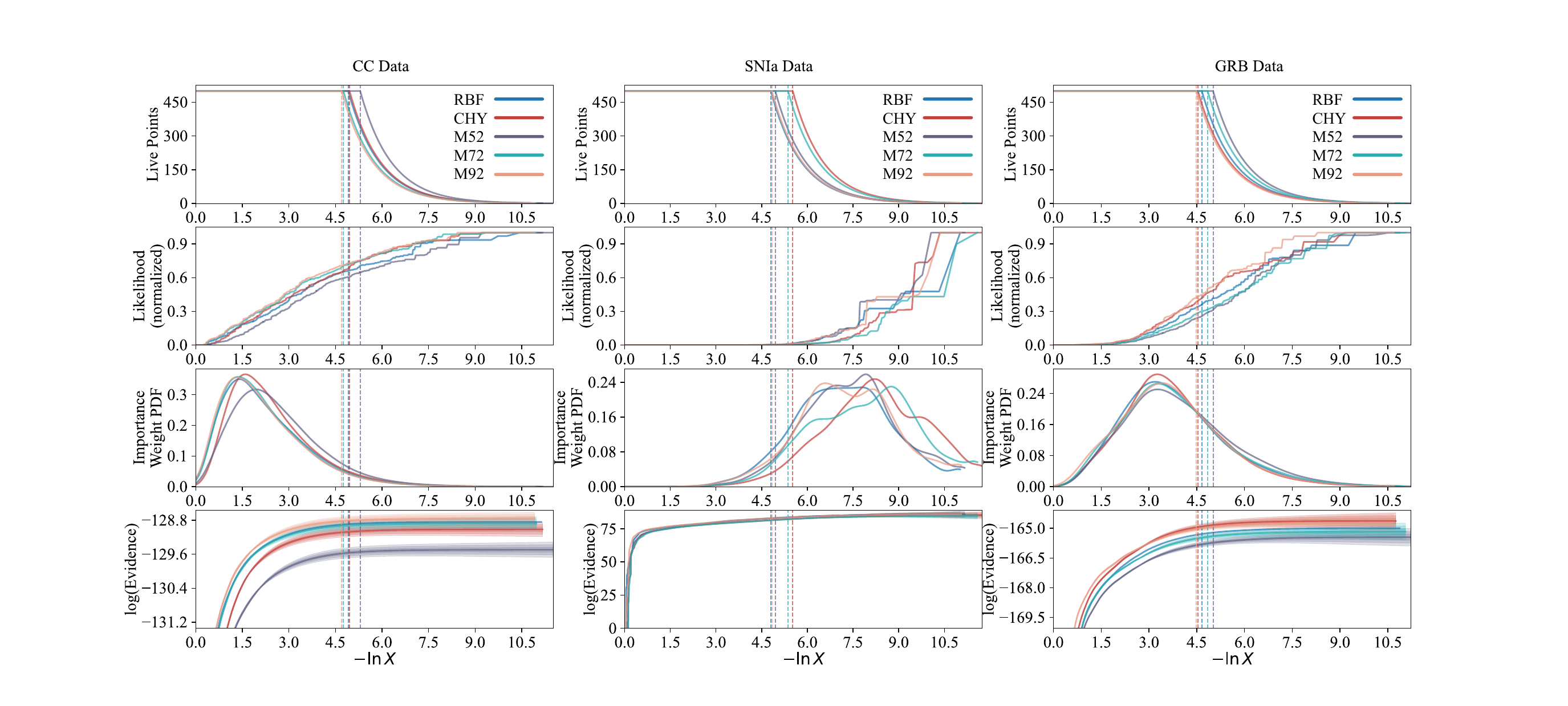}
\end{center}
\caption{Plots for sampler's running of 3 datasets. The horizontal axis of each subplot represents the natural logarithm of the a prior volume, $-\ln X$. The plot with the vertical axis of $\ln \mathcal{Z}$ is the plot we need for model comparison, and the value that tends to be stable is the model's corresponding evidence $\ln \mathcal{Z}$.}
\label{dypool}
\end{figure*}

The nested sampler of \verb"Dynesty" is utilized in this study to produce Figure \ref{dypool}. For each type of data, four figures are generated by the sampler. The bottom graph of each column in the figure is particularly useful for model selection, as it displays how the evidence for the model changes over the sampling process until it stabilizes. The other three plots in each column have different meanings, which are more thoroughly explained in \cite{speagle2020dynesty} and are not discussed in detail in our work.

\begin{figure*}[htbp]
\begin{center}
\includegraphics[scale=0.4]{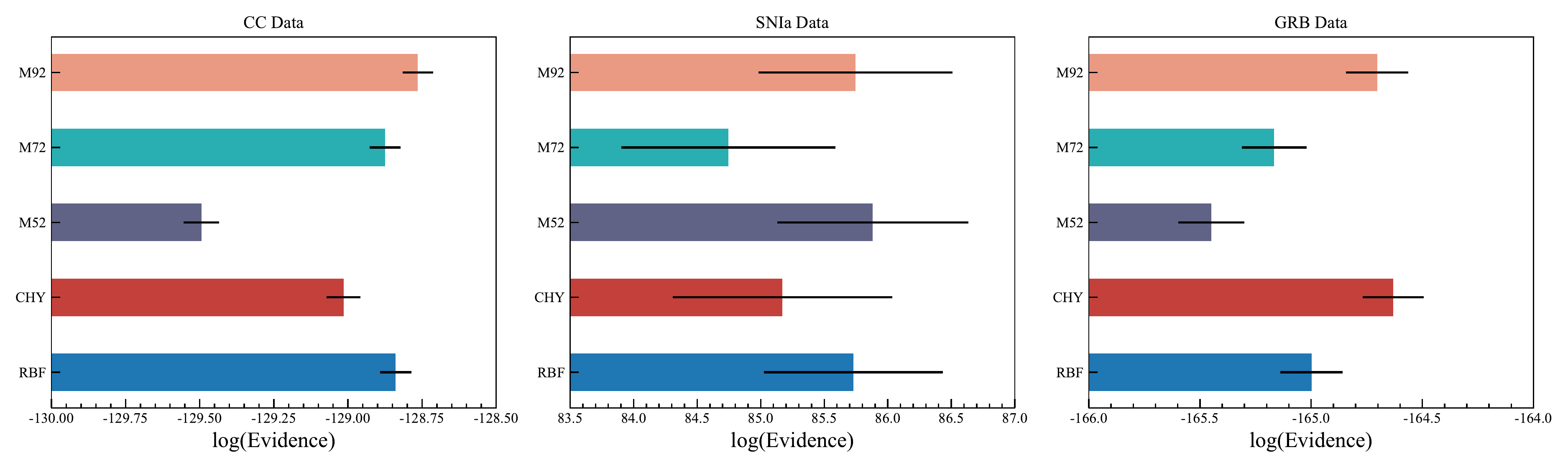}
\end{center}
\caption{Visualization plots of evidence in all cases. The solid black line indicates the error bar. The errors are consistent with those given in Table \ref{lnZfornest}. Note that for the CC and GRB data, the log evidence for them is negative.}
\label{evibar}
\end{figure*}

Remarkably, our results differ from those of ABC Rejection. In Figure \ref{dypool}, it is evident that M52 no longer dominates and even exhibits the lowest evidence for the CC data. The SNIa data, on the other hand, does not present an obvious result from the graph. To compare the results, we extract the exact runs' evidence results and display them in Table \ref{lnZfornest}. We then visualize these results using the bar chart in Figure \ref{evibar}.

\begin{deluxetable}{ccc}
\caption{The evidence $\ln\mathcal{Z}$ obtained for all types of data and kernels}\label{lnZfornest}
\tablewidth{0pt}
\tablehead{
\colhead{Data} & \colhead{Kernels} & \colhead{$\ln\mathcal{Z}$}
}
\startdata
     &RBF&-128.839$\pm$0.053\\
     &CHY&-129.015$\pm$0.057\\
CC   &M52&-129.494$\pm$0.060\\
     &M72&-128.875$\pm$0.052\\
     &M92&-128.764$\pm$0.051\\
\hline
     &RBF&85.729$\pm$0.705\\
     &CHY&85.171$\pm$0.863\\
SNIa &M52&85.882$\pm$0.751\\
     &M72&84.745$\pm$0.843\\
     &M92&85.746$\pm$0.762\\
\hline
     &RBF&-164.998$\pm$0.140\\
     &CHY&-164.631$\pm$0.138\\
GRB  &M52&-165.449$\pm$0.149\\
     &M72&-165.166$\pm$0.145\\
     &M92&-164.703$\pm$0.140\\
\enddata
\end{deluxetable}

From Figure \ref{evibar}, it is apparent that, except for the SNIa data where M52 is the best kernel, the other two cases show the opposite result, i.e. M52 is the ``worst" kernel. Even for the SNIa data, M52 does not display a decisive advantage over the other kernels.

\begin{figure*}[htbp]
\begin{center}
\includegraphics[scale=0.4]{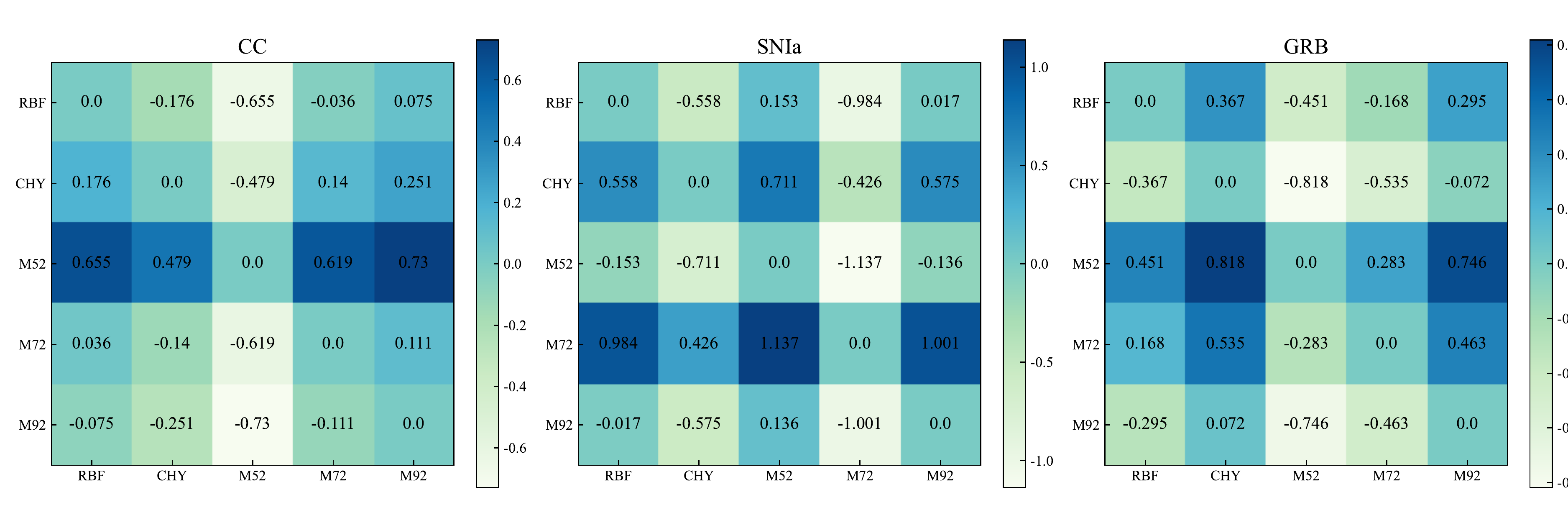}
\end{center}
\caption{Bayes factor heat map obtained by direct computation of the evidence. The meaning of each color block is the same as in Figure \ref{Bayefac}. As an example, for the color block in the second row and first column, the Bayes factor it represents is obtained from $\ln\mathcal{Z}_{\text{RBF}}-$ $\ln\mathcal{Z}_{\text{CHY}}$.}
\label{lnZbayes}
\end{figure*}

It is also important to understand how the nested sampling results reflect the Bayes factor's strength of evidence between different kernels, as discussed in Equation \eqref{lnz}. To display the Bayes factors between the kernels, similar to the approach used in ABC Rejection (Figure \ref{Bayefac}), we also generated a heatmap in Figure \ref{lnZbayes}.

To make the Bayes factor size associated with each color block identifiable, we label them in Figure \ref{lnZbayes}. However, since the difference between each kernel is minor, the Bayes factors do not display any significant differences. The largest Bayes factor only reaches the level of ``weak evidence" (M92 vs. M72 for SNIa data). Nearly all Bayes factors remain at the ``inconclusive" level, as shown in Table \ref{tab:lnK}. Therefore, we can assume that there is no significant difference among commonly used kernels in GP. Researchers can select kernel functions for their practical studies without concern for significant deviations in the results.

\section{conclusion} \label{sec:conclusion}
In this study, we apply ABC Rejection to investigate the topic of kernel selection for Gaussian processes (GP) using three alternative distance functions. These distance functions correspond to different ways of particle filtering using the ABC approach. Our results demonstrate that even without the additional sampling method of lowered thresholding, reliable kernel selection outcomes may be obtained. We find that the M52 kernel performs as the optimal kernel in all nine scenarios studied. Additionally, the widely employed RBF kernel function by astronomers does not perform as well as the other kernels, especially when compared to M52. A recent paper \citep{bernardo2021towards} reached a similar conclusion, but their analysis was limited to CC and SNIa data only and did not include a comparison of M52 to other kernels. Our research is more comprehensive and thorough, yielding more definitive results.

Moreover, the Bayes factor calculation results indicate that M52 outperforms the other options. The results reveal that, according to Table \ref{tab:K}, M52 can only attain a ``strong" level of evidence for the observed data in a few circumstances when compared to the RBF kernel. However, M52 does not exhibit very strong support for the observed data when compared to the other options. As a result, while M52 is the best kernel function in our research range, there is little statistical difference in the outputs produced by using kernels other than the RBF kernel.

On the other hand, we use a nested sampling algorithm and the \verb|Dynesty| package to directly evaluate the evidence for each kernel function to compare their performance under different data. This approach leads to different results from the ABC approach, where M52 no longer maintains its dominance in the evidence comparison, but instead becomes the kernel function with the lowest evidence in the CC and GRB data. However, this does not significantly affect our conclusions. By taking the log Bayes factors between them, we find that these differences are not remarkable, and no kernel function is appreciably supported by the data. Thus, based on the results of nested sampling, it does not matter which of these commonly used kernel functions a researcher chooses when using GP (although this conclusion is specific to the type of dataset we investigated).

There are several other interesting points to consider regarding nested sampling. First, the computational speed of the program is a point of interest for improvement. The slow speed of the program using the \verb|Dynesty| package makes it challenging to run the algorithm multiple times to rule out randomness in the results. In our work, the ABC Rejection algorithm runs a million times for each kernel, and the final result is obtained by averaging over bins (100 times for a bin). However, repeated runs of this size cannot be performed for nested sampling. Second, we find that although repeated nested sampling leads to small fluctuations in the evidence for each kernel function, for both CC and GRB data, M52 is always the kernel function with the lowest evidence, and the results vary from time to time as to which one has the highest evidence. In contrast, algorithms such as ABC Rejection and similar ABC-SMC \citep{bernardo2021towards} give the exact opposite results. Although this does not affect the final selection, this result, which is unlikely to be the result of statistical error, deserves further investigation.

As a future direction of this study, there are potential improvements to be made in the approach of finding posterior distributions. In addition to ABC Rejection, the posterior distribution can also be generated using VBMC\citep{acerbi2018variational}. This is achieved by progressively modifying a given distribution to resemble the true distribution based on the observed data. To further limit the selection of kernel, the introduction of more simulated data to populate the observed data is possible\citep{bernstein2012supernova}. Another approach that could be considered is the use of Sparse Markovian Gaussian processes\citep{wilkinson2021sparse}. This method utilizes several local Gaussian terms to approximate a non-Gaussian likelihood function, making it particularly suitable for large sample data. As the likelihood function of cosmological data is also uncertain, the issue of kernel selection for Sparse Markovian Gaussian processes is of great significance.

We express our gratitude to the anonymous reviewer for the constructive comments, and to Bo Yu, Xu-Yao Gong, Jing Niu, Yuan-Hang Mao, Yuan-Bo Xie, and Kang Jiao for their insightful discussions. We would also like to acknowledge the contributions of the editors. The python packages \verb|NumPy|\citep{harris2020array} and \verb|Matplotlib|\citep{Hunter:2007} played a critical role in our research, and we extend our thanks to their developers. This work is supported by grants from the National Science Foundation of China (Nos.11929301 and 61802428)


\bibliography{kselect}{}

\begin{thebibliography}{}
\expandafter\ifx\csname natexlab\endcsname\relax\def\natexlab#1{#1}\fi
\providecommand{\url}[1]{\href{#1}{#1}}
\providecommand{\dodoi}[1]{doi:~\href{http://doi.org/#1}{\nolinkurl{#1}}}
\providecommand{\doeprint}[1]{\href{http://ascl.net/#1}{\nolinkurl{http://ascl.net/#1}}}
\providecommand{\doarXiv}[1]{\href{https://arxiv.org/abs/#1}{\nolinkurl{https://arxiv.org/abs/#1}}}

\bibitem[{Abdessalem {et~al.}(2017)Abdessalem, Dervilis, Wagg, \&
  Worden}]{abdessalem2017automatic}
Abdessalem, A.~B., Dervilis, N., Wagg, D.~J., \& Worden, K. 2017, Frontiers in
  Built Environment, 3, 52

\bibitem[{Acerbi(2018)}]{acerbi2018variational}
Acerbi, L. 2018, Advances in Neural Information Processing Systems, 31

\bibitem[{Akeret {et~al.}(2015)Akeret, Refregier, Amara, Seehars, \&
  Hasner}]{akeret2015approximate}
Akeret, J., Refregier, A., Amara, A., Seehars, S., \& Hasner, C. 2015, Journal
  of Cosmology and Astroparticle Physics, 2015, 043

\bibitem[{Beaumont {et~al.}(2009)Beaumont, Cornuet, Marin, \&
  Robert}]{beaumont2009adaptive}
Beaumont, M.~A., Cornuet, J.-M., Marin, J.-M., \& Robert, C.~P. 2009,
  Biometrika, 96, 983

\bibitem[{Bernardo \& Said(2021)}]{bernardo2021towards}
Bernardo, R.~C., \& Said, J.~L. 2021, Journal of Cosmology and Astroparticle
  Physics, 2021, 027

\bibitem[{Bernstein {et~al.}(2012)Bernstein, Kessler, Kuhlmann, Biswas, Kovacs,
  Aldering, Crane, D'andrea, Finley, Frieman,
  {et~al.}}]{bernstein2012supernova}
Bernstein, J., Kessler, R., Kuhlmann, S., {et~al.} 2012, The Astrophysical
  Journal, 753, 152

\bibitem[{Bonassi \& West(2015)}]{bonassi2015sequential}
Bonassi, F.~V., \& West, M. 2015, Bayesian Analysis, 10, 171

\bibitem[{Capp{\'e} {et~al.}(2004)Capp{\'e}, Guillin, Marin, \&
  Robert}]{cappe2004population}
Capp{\'e}, O., Guillin, A., Marin, J.-M., \& Robert, C.~P. 2004, Journal of
  Computational and Graphical Statistics, 13, 907

\bibitem[{Demianski {et~al.}(2017)Demianski, Piedipalumbo, Sawant, \&
  Amati}]{demianski2017cosmology}
Demianski, M., Piedipalumbo, E., Sawant, D., \& Amati, L. 2017, Astronomy \&
  Astrophysics, 598, A112

\bibitem[{Dhawan {et~al.}(2021)Dhawan, Alsing, \& Vagnozzi}]{dhawan2021non}
Dhawan, S., Alsing, J., \& Vagnozzi, S. 2021, Monthly Notices of the Royal
  Astronomical Society: Letters, 506, L1

\bibitem[{Feroz {et~al.}(2009)Feroz, Hobson, \& Bridges}]{feroz2009multinest}
Feroz, F., Hobson, M., \& Bridges, M. 2009, Monthly Notices of the Royal
  Astronomical Society, 398, 1601

\bibitem[{G{\'o}mez-Valent \& Amendola(2018)}]{gomez2018h0}
G{\'o}mez-Valent, A., \& Amendola, L. 2018, Journal of Cosmology and
  Astroparticle Physics, 2018, 051

\bibitem[{Handley {et~al.}(2015{\natexlab{a}})Handley, Hobson, \&
  Lasenby}]{handley2015polychord}
Handley, W., Hobson, M., \& Lasenby, A. 2015{\natexlab{a}}, Monthly Notices of
  the Royal Astronomical Society, 453, 4384

\bibitem[{Handley {et~al.}(2015{\natexlab{b}})Handley, Hobson, \&
  Lasenby}]{handley2015polychord_2}
---. 2015{\natexlab{b}}, Monthly Notices of the Royal Astronomical Society:
  Letters, 450, L61

\bibitem[{Harris {et~al.}(2020)Harris, Millman, van~der Walt, Gommers,
  Virtanen, Cournapeau, Wieser, Taylor, Berg, Smith, Kern, Picus, Hoyer, van
  Kerkwijk, Brett, Haldane, del R{\'{i}}o, Wiebe, Peterson,
  G{\'{e}}rard-Marchant, Sheppard, Reddy, Weckesser, Abbasi, Gohlke, \&
  Oliphant}]{harris2020array}
Harris, C.~R., Millman, K.~J., van~der Walt, S.~J., {et~al.} 2020, Nature, 585,
  357, \dodoi{10.1038/s41586-020-2649-2}

\bibitem[{Hunter(2007)}]{Hunter:2007}
Hunter, J.~D. 2007, Computing in Science \& Engineering, 9, 90,
  \dodoi{10.1109/MCSE.2007.55}

\bibitem[{Hwang {et~al.}(2023)Hwang, L'Huillier, Keeley, Jee, \&
  Shafieloo}]{hwang2023use}
Hwang, S.-g., L'Huillier, B., Keeley, R.~E., Jee, M.~J., \& Shafieloo, A. 2023,
  Journal of Cosmology and Astroparticle Physics, 2023, 014

\bibitem[{Ishida {et~al.}(2015)Ishida, Vitenti, Penna-Lima, Cisewski, de~Souza,
  Trindade, Cameron, Busti, collaboration, {et~al.}}]{ishida2015cosmoabc}
Ishida, E.~E., Vitenti, S.~D., Penna-Lima, M., {et~al.} 2015, Astronomy and
  Computing, 13, 1

\bibitem[{Jeffreys(1998)}]{jeffreys1998theory}
Jeffreys, H. 1998, The theory of probability (OUP Oxford)

\bibitem[{Jennings \& Madigan(2017)}]{jennings2017astroabc}
Jennings, E., \& Madigan, M. 2017, Astronomy and computing, 19, 16

\bibitem[{Jiao {et~al.}(2022)Jiao, Borghi, Moresco, \& Zhang}]{Jiao2022new}
Jiao, K., Borghi, N., Moresco, M., \& Zhang, T.-J. 2022, arXiv preprint
  arXiv:2205.05701

\bibitem[{Jimenez {et~al.}(2003)Jimenez, Verde, Treu, \& Stern}]{Jimenez_2003}
Jimenez, R., Verde, L., Treu, T., \& Stern, D. 2003, The Astrophysical Journal,
  593, 622, \dodoi{10.1086/376595}

\bibitem[{Kilbinger {et~al.}(2012)Kilbinger, Benabed, Capp{\'e}, Coupon,
  Cardoso, Fort, McCracken, Prunet, Robert, \& Wraith}]{kilbinger2012cosmopmc}
Kilbinger, M., Benabed, K., Capp{\'e}, O., {et~al.} 2012, Astrophysics Source
  Code Library, ascl

\bibitem[{Mehrabi \& Basilakos(2020)}]{mehrabi2020does}
Mehrabi, A., \& Basilakos, S. 2020, The European Physical Journal C, 80, 1

\bibitem[{Moresco(2015)}]{10.1093/mnrasl/slv037}
Moresco, M. 2015, Monthly Notices of the Royal Astronomical Society: Letters,
  450, L16, \dodoi{10.1093/mnrasl/slv037}

\bibitem[{Moresco {et~al.}(2012)Moresco, Cimatti, Jimenez, Pozzetti, Zamorani,
  Bolzonella, Dunlop, Lamareille, Mignoli, Pearce, Rosati, Stern, Verde, Zucca,
  Carollo, Contini, Kneib, F{\`{e}}vre, Lilly, Mainieri, Renzini, Scodeggio,
  Balestra, Gobat, McLure, Bardelli, Bongiorno, Caputi, Cucciati, de~la Torre,
  de~Ravel, Franzetti, Garilli, Iovino, Kampczyk, Knobel, Kova{\v{c}}, Borgne,
  Brun, Maier, Pell{\'{o}}, Peng, Perez-Montero, Presotto, Silverman, Tanaka,
  Tasca, Tresse, Vergani, Almaini, Barnes, Bordoloi, Bradshaw, Cappi, Chuter,
  Cirasuolo, Coppa, Diener, Foucaud, Hartley, Kamionkowski, Koekemoer,
  L{\'{o}}pez-Sanjuan, McCracken, Nair, Oesch, Stanford, \&
  Welikala}]{Moresco_2012}
Moresco, M., Cimatti, A., Jimenez, R., {et~al.} 2012, Journal of Cosmology and
  Astroparticle Physics, 2012, 006, \dodoi{10.1088/1475-7516/2012/08/006}

\bibitem[{Moresco {et~al.}(2016)Moresco, Pozzetti, Cimatti, Jimenez, Maraston,
  Verde, Thomas, Citro, Tojeiro, \& Wilkinson}]{Moresco_2016}
Moresco, M., Pozzetti, L., Cimatti, A., {et~al.} 2016, Journal of Cosmology and
  Astroparticle Physics, 2016, 014, \dodoi{10.1088/1475-7516/2016/05/014}

\bibitem[{Morey {et~al.}(2016)Morey, Romeijn, \& Rouder}]{MOREY20166}
Morey, R.~D., Romeijn, J.-W., \& Rouder, J.~N. 2016, Journal of Mathematical
  Psychology, 72, 6, \dodoi{https://doi.org/10.1016/j.jmp.2015.11.001}

\bibitem[{Mukherjee \& Banerjee(2021)}]{mukherjee2021non}
Mukherjee, P., \& Banerjee, N. 2021, The European Physical Journal C, 81, 1

\bibitem[{Ratsimbazafy {et~al.}(2017)Ratsimbazafy, Loubser, Crawford, Cress,
  Bassett, Nichol, \& Väisänen}]{10.1093/mnras/stx301}
Ratsimbazafy, A.~L., Loubser, S.~I., Crawford, S.~M., {et~al.} 2017, Monthly
  Notices of the Royal Astronomical Society, 467, 3239,
  \dodoi{10.1093/mnras/stx301}

\bibitem[{Reichart {et~al.}(2001)Reichart, Lamb, Fenimore, Ramirez-Ruiz, Cline,
  \& Hurley}]{reichart2001possible}
Reichart, D.~E., Lamb, D.~Q., Fenimore, E.~E., {et~al.} 2001, The Astrophysical
  Journal, 552, 57

\bibitem[{Sarro {et~al.}(2012)Sarro, Eyer, O'Mullane, \&
  De~Ridder}]{sarro2012astrostatistics}
Sarro, L.~M., Eyer, L., O'Mullane, W., \& De~Ridder, J. 2012, Astrostatistics
  and Data Mining, Vol.~2 (Springer Science \& Business Media)

\bibitem[{Scolnic {et~al.}(2018)Scolnic, Jones, Rest, Pan, Chornock, Foley,
  Huber, Kessler, Narayan, Riess, {et~al.}}]{scolnic2018complete}
Scolnic, D.~M., Jones, D., Rest, A., {et~al.} 2018, The Astrophysical Journal,
  859, 101

\bibitem[{Seikel \& Clarkson(2013)}]{seikel2013optimising}
Seikel, M., \& Clarkson, C. 2013, arXiv preprint arXiv:1311.6678

\bibitem[{Seikel {et~al.}(2012)Seikel, Clarkson, \&
  Smith}]{seikel2012reconstruction}
Seikel, M., Clarkson, C., \& Smith, M. 2012, Journal of Cosmology and
  Astroparticle Physics, 2012, 036

\bibitem[{Shafieloo {et~al.}(2012)Shafieloo, Kim, \&
  Linder}]{PhysRevD.85.123530}
Shafieloo, A., Kim, A.~G., \& Linder, E.~V. 2012, Phys. Rev. D, 85, 123530,
  \dodoi{10.1103/PhysRevD.85.123530}

\bibitem[{Simon {et~al.}(2005)Simon, Verde, \& Jimenez}]{PhysRevD.71.123001}
Simon, J., Verde, L., \& Jimenez, R. 2005, Phys. Rev. D, 71, 123001,
  \dodoi{10.1103/PhysRevD.71.123001}

\bibitem[{Skilling(2006)}]{10.1214/06-BA127}
Skilling, J. 2006, Bayesian Analysis, 1, 833 , \dodoi{10.1214/06-BA127}

\bibitem[{Speagle(2020)}]{speagle2020dynesty}
Speagle, J.~S. 2020, Monthly Notices of the Royal Astronomical Society, 493,
  3132

\bibitem[{Stern {et~al.}(2010)Stern, Jimenez, Verde, Kamionkowski, \&
  Stanford}]{Stern_2010}
Stern, D., Jimenez, R., Verde, L., Kamionkowski, M., \& Stanford, S.~A. 2010,
  Journal of Cosmology and Astroparticle Physics, 2010, 008,
  \dodoi{10.1088/1475-7516/2010/02/008}

\bibitem[{Sun {et~al.}(2021)Sun, Jiao, \& Zhang}]{sun2021influence}
Sun, W., Jiao, K., \& Zhang, T.-J. 2021, The Astrophysical Journal, 915, 123

\bibitem[{Toni \& Stumpf(2009)}]{toni2009tutorial}
Toni, T., \& Stumpf, M.~P. 2009, arXiv preprint arXiv:0910.4472

\bibitem[{Toni \& Stumpf(2010)}]{toni2010simulation}
---. 2010, Bioinformatics, 26, 104

\bibitem[{Trotta(2008)}]{trotta2008bayes}
Trotta, R. 2008, Contemporary Physics, 49, 71

\bibitem[{Turner \& Van~Zandt(2012)}]{turner2012tutorial}
Turner, B.~M., \& Van~Zandt, T. 2012, Journal of Mathematical Psychology, 56,
  69

\bibitem[{Weyant {et~al.}(2013)Weyant, Schafer, \&
  Wood-Vasey}]{weyant2013likelihood}
Weyant, A., Schafer, C., \& Wood-Vasey, W.~M. 2013, The Astrophysical Journal,
  764, 116

\bibitem[{Wilkinson {et~al.}(2021)Wilkinson, Solin, \&
  Adam}]{wilkinson2021sparse}
Wilkinson, W., Solin, A., \& Adam, V. 2021, in International Conference on
  Artificial Intelligence and Statistics, PMLR, 1747--1755

\bibitem[{Williams \& Rasmussen(2006)}]{williams2006gaussian}
Williams, C.~K., \& Rasmussen, C.~E. 2006, Gaussian processes for machine
  learning, Vol.~2 (MIT press Cambridge, MA)

\bibitem[{Yang {et~al.}(2013)Yang, Yu, Zhang, \& Zhang}]{yang2013improved}
Yang, X., Yu, H.-R., Zhang, Z.-S., \& Zhang, T.-J. 2013, The Astrophysical
  Journal Letters, 777, L24

\bibitem[{Zhang {et~al.}(2014)Zhang, Zhang, Yuan, Liu, Zhang, \&
  Sun}]{Zhang_2014}
Zhang, C., Zhang, H., Yuan, S., {et~al.} 2014, Research in Astronomy and
  Astrophysics, 14, 1221, \dodoi{10.1088/1674-4527/14/10/002}

\bibitem[{Zuntz {et~al.}(2015)Zuntz, Paterno, Jennings, Rudd, Manzotti,
  Dodelson, Bridle, Sehrish, \& Kowalkowski}]{zuntz2015cosmosis}
Zuntz, J., Paterno, M., Jennings, E., {et~al.} 2015, Astronomy and Computing,
  12, 45

\bibitem[{Ó~Colgáin \& Sheikh-Jabbari(2021)}]{o_colgain_elucidating_2021}
Ó~Colgáin, E., \& Sheikh-Jabbari, M.~M. 2021, The European Physical Journal
  C, 81, 892, \dodoi{10.1140/epjc/s10052-021-09708-2}

\end{thebibliography}
\bibliographystyle{aasjournal}


\end{document}